# Reliable Editions from Unreliable Components: Estimating Ebooks from Print Editions Using Profile Hidden Markov Models

A. B. RIDDELL, Indiana University Bloomington, United States

A profile hidden Markov model, a popular model in biological sequence analysis, can be used to model related sequences of characters transcribed from books, magazines, and other printed materials. This paper documents one application of a profile HMM: automatically producing an ebook edition from distinct print editions. The resulting ebook has virtually all the desired properties found in a publisher-prepared ebook, including accurate transcription and an absence of print artifacts such as end-of-line hyphenation and running headers. The technique, which has particular benefits for readers and libraries that require books in an accessible format, is demonstrated using seven copies of a nineteenth-century novel.

## 1 INTRODUCTION

When publishers bring a book to market today, they offer two versions for sale, a print version and an ebook version. In most cases, a reader encountering a print version and an ebook version at the same time will notice few differences. The sequence of characters visible on the page and on the e-reader will be identical, save for running headers, page numbers, end-of-line hyphenation, and page breaks—features exclusively found in print versions. Although the sequence of characters in each version is virtually identical, the two versions differ in significant ways. As ebooks abide in computer memory, the ebook version costs almost nothing to store, transport, or reproduce. Ebooks are also markedly more accessible to readers with blindness and print disabilities as they can be read using text-to-speech and Braille devices. Readers using ebooks can also improve the legibility of a text by increasing the font size.

Producing an ebook and a print version of a book is laborious and costly. Publishers cannot automatically create one version from the other. Producing a version from a manuscript that can be printed and bound by a press requires the labor of a book designer. Producing the ebook version requires the labor of an ebook designer.

Nonprofit and low-profit publishers would therefore welcome a technique for automatically creating an ebook version from a print version. Readers and libraries seeking ebook editions of older books also stand to benefit from such a technique: any tool that can "recover" an ebook version from a print version can be mechanically applied to existing library books that lack ebook versions.

This paper describes how to automatically estimate an ebook from several print copies using a profile hidden Markov model (profile HMM), a model introduced in biological sequence analysis in 1993 [5, 8, 11]. Unlike texts produced from page images via optical character recognition (OCR), ebooks estimated using the method described here are potentially indistinguishable from publisher-prepared ebooks. I demonstrate the use of a profile HMM to estimate an ebook version of a chapter of Charles Dickens' *David Copperfield* (1850) using seven print copies of the novel. The performance of this method is evaluated by calculating the percentage of matching characters in an alignment between the estimated ebook version and a reference ebook. The profile HMM produces an ebook version that aligns as well with the reference ebook as the Project Gutenberg ebook, a version on which human editors labored for tens if not hundreds of hours.

Author's address: A. B. Riddell, Indiana University Bloomington, United States, riddella@indiana.edu.

## 2  COMPARISON OF TRANSCRIBED BOOK PAGES AND AN EBOOK

Here I introduce terminology and give a concrete example of the differences between transcribed book pages and the machine-readable contents of an ebook. In early 2022, these preliminary notes seem necessary. The dominant multinational publishing companies have only recently embraced the practice of simultaneously publishing print and ebook versions of new works. Moreover, during the early years of ebook publishing a variety of technical standards circulated. It is only very recently (2018) that a standard, EPUB 3, anchored in the HTML5 and CSS3 standards used by web browsers gained widespread adoption.

In October 2020, the London-based publisher Verso made *Culture and Materialism* by Raymond Williams available for purchase in two formats: paperback (ISBN 9781788738606) and ebook (ISBN 9781789600049).[1] The contents of these two formats, if suitably prepared, can be compared. The paperback version must first be transcribed into a machine-readable form. That is, for each discrete mark on each page of the paperback book, a counterpart must be chosen from the ca. 140,000 characters in the Unicode Standard. For example, a mark resembling "a" would likely be recorded using the character "a", U+0061 ("Latin Small Letter A"). Horizontal space, line breaks, and page breaks are transcribed using standard conventions: U+0020, U+000A, and U+000C, respectively. Marks on the page, including illustrations, which are not described in the Unicode Standard are ignored. Transcriptions of a page will vary because human and machine transcribers make inadvertent mistakes and enjoy freedom to choose among alternatives. For example, a machine or human may accidentally transcribe a mark resembling "A" as "H", or as one of numerous other characters, such as "А", U+0410 ("Cyrillic Capital Letter A"). Less work is required to prepare the ebook version for comparison. Because a computer file using the EPUB 3 format consist of files stored in a container (ZIP), the ebook must be uncompressed to get at the enclosed XHTML content documents. (In the case of a book published by Verso this step is straightforward because Verso distributes EPUB files unencumbered by encryption.)

To facilitate comparison among ebook versions of similar works and ebooks of the same work in different formats (e.g., EPUB 2, Mobipocket, HTML, CHM), the XHTML content documents are mechanically converted to HTML5 and stripped of redundant markup and whitespace. At this point we have two sequences of characters that can be compared. I will refer to a machine-readable transcribed sequence as a "print sequence". I will refer to an HTML5 sequence extracted from an ebook as an "ebook sequence".

If we align a print sequence obtained via OCR and an ebook sequence of *Culture and Materialism*, we find considerable agreement. Characters in the two sequences match most of the time and virtually all of the differences are predictable. Relative to the ebook sequence, the print sequence inserts running headers, page numbers, end-of-line hyphenation, line breaks, and page breaks. Relative to the print sequence, the ebook sequence inserts HTML tags indicating paragraph breaks (<p>) and emphasis (<em>). That the two sequences align as well as they do is perhaps unsurprising. A stated aim of the Unicode Standard is to facilitate transcription of printed books and articles [3, p. 1].

We are not limited to considering pairwise alignment between sequences. Multiple sequences may also be aligned and compared. For example, we could compare the ebook sequence and the print sequence with sequences derived from the five previous paperback and hardback editions of William's work.[2]

---

[1] In early 2022, both versions remain available for purchase on the publisher's website, https://www.versobooks.com/books/43-culture-and-materialism.
[2] Williams, R. (1980). *Problems in Materialism and Culture.* London: Verso. ISBN:086091-028-8 (Hardback); Williams, R. (1980). *Problems in Materialism and Culture.* London: Verso. ISBN:086091-729-0 (Paperback); Williams, R. (1997). *Problems in Materialism and Culture.* London: Verso. ISBN:185984-113-9 (Paperback); Williams, R. (2006). *Culture and Materialism.* London: Verso. ISBN: 978-1-84467-060-4 (Paperback); Williams, R. (2010). *Culture and Materialism.* London: Verso. ISBN: 978-1-84467-663-7 (Hardback).

```
HBA_HUMAN     ...VGA--HAGEY...
HBB_HUMAN     ...V----NVDEV...
MYG_PHYCA     ...VEA--DVAGH...
GLB3_CHITP    ...VKG------D...
GLB5_PETMA    ...VYS--TYETS...
LGB2_LUPLU    ...FNA--NIPKH...
GLB1_GLYDI    ...IAGADNGAGV...
```

Fig. 1. Part of a multiple alignment of seven globin proteins from Durbin et al. [4, p. 107]. Letters correspond to amino acids. As eaxmples, "V" is valine, "G" is glcyine, "A" is Alanine. A "-" indicates either a deletion in a particular sequence or occupies space where another sequence has an idiosyncratic insertion. The left hand column names the proteins. For example, "HBA_HUMAN" is human hemoglobin A, one piece of the human hemoglobin tetramer.

## 3 PROFILE HIDDEN MARKOV MODEL

To estimate an ebook sequence, homologous print sequences are collected and then used to estimate a profile hidden Markov model (profile HMM). (*Homology* is similarity owing to common descent.) The fitted profile HMM's most probable sequence—its mode or modal sequence—is then used as the estimated ebook sequence. To my knowledge, this is the first time the model has been used to model character sequences transcribed from print materials.

The profile HMM has been a staple of biological sequence analysis since 1993 [5]. The procedure described here is a trivial extension of prior art.

In this section I review the profile HMM for an audience familiar with hidden Markov models. Then I review standard methods for estimating a profile HMM from observed sequences.

### 3.1 Background and model architecture

The profile HMM provides a probabilistic approach to the problem of multiple alignment. In the context in which it was introduced, the sequences being aligned were proteins and RNAs. (The "profile" in the model's name references earlier methods in biological sequence analysis.) A typical use of a profile HMM in biology is identifying potentially homologous proteins, such as oxygen-carrying (globin) poteins in different mammals. These proteins—sequences of amino acids—tend to resemble each other. They are, however, far from identical. A profile HMM provides the needed method for characterizing differences and commonalities among sequences. Parameters of a profile HMM are typically initially estimated using a small selection of sequences already known to be homologous. The fitted profile HMM is then used in other tasks, such as searching for additional related proteins or evaluating the plausibility of a particular multiple alignment.

The architecture of the profile HMM is best understood by reference to a multiple alignment, such as the one shown in Figure 1. The profile HMM is an inhomogeneous HMM—its hidden states are position-specific. In essence, each state is associated with a column of the multiple alignment. Amino acid residues (or, in our case, characters) associated with consensus columns are emitted by *match* states. *Insert* states at each position allow for the possibility that a particular sequence may have one or more idiosyncratic insertions relative to the sequence of *match* states. *Delete* states allow for a particular sequence to lack a residue (character) at a position where other sequences tend to have one. *Delete* states do not emit a residue (character)—they are "silent states".

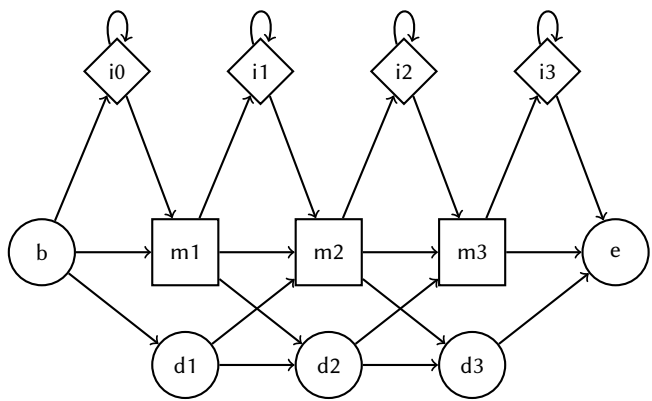

Fig. 2. A profile HMM with three match states (m1, m2, m3).

Transitions between states are constrained to flow in one direction, as arrows in Figure 2 indicate. A *match* or *delete* state, once visited, is never re-visited. An *insert* state can be revisited, but only from the *insert* state itself. Dwelling in an *insert* state allows for a series of idiosyncratic residues (characters) to be emitted—e.g., a running header found in only one print sequence.

Figure 2 shows the profile HMM architecture for a model with three *match* states. Note the presence of three additional special states (labeled "b", "i0", and "e") that are used to model the beginning and ending of an observed sequence. When calculating the probability of a sequence, a profile HMM always begins in a silent *begin* state and ends in a silent *end* state. The first *insert* state ("i0") allows for the observation of idiosyncratic insertions before observing an emission associated with the first *match* state.

Beyond expanding the alphabet to accommodate characters that occur in print sequences, no modification of the profile HMM is required to use it in the context of print sequences. Just as one might estimate a profile HMM using seven hemoglobin A proteins from seven different mammals (length ≈ 140 residues), one can estimate a profile HMM using seven different copies of a 19th-century novel (length ≈ 1,975,000 characters). There are no conceptual differences between the two models. Both can be analyzed and used in precisely the same way.

The most important thing to appreciate about a profile HMM is that it is not a model of any particular sequence. A profile HMM describes a distribution over sequences. Although the profile HMM is designed such that it assigns high probability to sequences similar to those used to estimate the model's parameters (typically homologous sequences), the model's support includes arbitrary sequences. A profile HMM will assign a probability to any sequence [4, p. 108]. This feature of the profile HMM makes it easy to use to search for (distantly) related sequences. Sequences in a database can be scored in terms of their probability under the model. Sequences that score conspicuously higher than others merit consideration as potential members of the family.

For those coming to profile HMMs from computer science or computational linguistics [9], a distinctive feature of the profile HMM is its many hidden states. As there are three hidden states associated with each position in the profile HMM—with each consensus column, essentially—the number of hidden states in a profile HMM is three times greater than the length of the typical sequence the model is designed for. By contrast, HMMs featured in textbook introductions have a handful of hidden states (often two) [4, Ch. 3][12][13][10, Appendix A].

The large number of states creates no theoretical issues but it does pose a computational problem. In general, when using the forward algorithm to calculate the probability of an observation of length $T$ in an HMM with $N$ hidden states, $TN^2$ calculations are required [9, p. 255]. In the case of the profile HMM, $T \approx N/3$. Naive software implementations of the profile HMM for character sequences will, as a rule, struggle to deal with character sequence families as long as a single chapter. Fortunately, a variety of techniques, typically involving approximations, exists to deal with this problem [2].

Those familiar with biological sequence analysis will note that there are other variants of the original profile HMM—a model introduced in Haussler et al. [8] and Krogh et al. [11]. It will do no harm to speak of "the profile HMM" in the present setting. In this paper, I only consider the problem of global alignment and "profile HMM" only refers to the original profile HMM or to its streamlined version, the "Plan 7" profile HMM introduced by Eddy [5]. The Plan 7 variant removes transitions from *insert* states to *delete* states and from *delete* states to *insert* states. When used for global alignment, the differences between the original Krogh/Haussler profile HMM and the Plan 7 profile HMM are slight and primarily of interest to those writing software.

## 3.2 Estimating a profile HMM's parameters

A profile HMM is an ordinary hidden Markov model whose parameters may be estimated using standard Bayesian methods. Typically maximum a posteriori parameter estimates are obtained using expectation maximization (aka Baum-Welch) or gradient ascent. (Gibbs sampling and Hamiltonian Monte Carlo (HMC) may also be used.) Whatever the method used, care must be taken to avoid numeric underflow and to choose suitable initial parameter estimates. The number of hidden states in the profile HMM—proportional to "model length" (the number of *match* states)—must also be picked in advance. In addition to presenting the profile HMM and its uses in biology, Durbin et al. [4] provide guidance on these choices and describes in detail how to implement the original Krogh/Haussler profile HMM.

The same procedures for fixing model length and choosing initial parameters can be used for character sequences as for amino acid sequences. Durbin et al. [4] make several recommendations. Model length can be set to be the average length of the sequences being studied or can be taken from an initial guess at the number of consensus columns based on a provisional multiple alignment using a dynamic programming algorithm [4, §6.4]. Initial parameters can be sampled from prior distributions or, again, derived from an initial guess at a multiple alignment [4, §6.5].

In the case of character sequences derived from books, using an initial guess at a multiple alignment to estimate an initial model may tend to work better than it does in the case of biological sequences because homologous character sequences tend to exhibit less variation. (Heuristic alignment algorithms work well when sequences being aligned exhibit low variation [4, p. 137].) Starting with parameters close to the maximum a posteriori parameter estimates reduces required computation time and helps avoid inferior local maxima.

Finally, there is the choice of model architecture. Using the original profile HMM instead of the streamlined Plan 7 variant is unlikely to affect the probability a model assigns to different alignments [4, §5.2]. The modal sequence of the fitted model, our chief concern here, is likely to be identical. There is no reason not to use the streamlined Plan 7 model architecture as it is slightly simpler, slightly easier to implement, and has fewer transition probability parameters that must be estimated. The Plan 7 architecture is also used in software in general use today [6].

Durbin et al. [4] describe how to implement the original Krogh/Haussler profile HMM. As the differences between the two models are minor and easy to understand, adjusting the provided implementation instructions for the newer architecture is straightforward.

## 4 EXPERIMENT: ESTIMATING AN EBOOK EDITION OF *DAVID COPPERFIELD*

In order to verify that the procedure works as expected, I estimate a profile HMM for a randomly-chosen chapter of *David Copperfield* using seven homologous print sequences. The model's modal sequence is then compared with a reference ebook version.

I estimate the parameters of the profile HMM by using statistics derived from a guess at a multiple alignment, following the recommendation of Durbin et al. [4, p. 153]. I use the non-probabilistic Barton–Sternberg iterative multiple alignment algorithm to arrive at the initial guess [4, p. 149–150]. I then use a simple heuristic to identify *match* states: any column that has less than 50% gap symbols is associated with a *match* state [4, p. 123]. Given a multiple alignment with *match* states marked, serviceable approximations of the maximum a posteriori estimates of all the profile HMM's parameters—the emission and transition probability distributions—can be obtained [4, p. 123-124].

The length of model (16,783) makes using Baum-Welch or gradient ascent prohibitively expensive for a naive implementation. So, for the moment, I do not perform any gradient ascent steps. I use the initial parameter estimates without any further fitting. More careful estimation of model parameters will certainly be required when only two different print copies are available as observations. Here, with seven copies, it seems unlikely to make a significant difference. (As Durbin et al. note, "[a]n alignment of very similar sequences will generally be unambiguous" [4, p. 137].) To check this claim, I estimate a smaller model of the initial paragraph (length 625) using the same seven copies, allowing gradient ascent to run for 10 epochs. No change to the modal sequence is observed.

As mentioned above, I use a 107-character alphabet for all emission distributions. This alphabet includes characters frequently found in publisher-prepared ebook editions and an opportunistic sample of Internet Archive print sequences: other *David Copperfield* chapters as well as chapters from print sequences from George Eliot's *The Mill on the Floss* (1860). When calculating empirical distributions from sequences that contain out-of-alphabet characters, such characters are replaced with spaces.

Software source code and data are available at https://doi.org/10.5281/zenodo.6412406.

### 4.1 Seven copies of *David Copperfield*

In this experiment, I work with seven copies of Charles Dickens' *David Copperfield* (1850). I use a single chapter—Chapter 29—selected uniformly at random, rather than the entire book to reduce computation time.

*David Copperfield* was selected for two reasons. First, the novel is in the public domain. Because public domain works can be freely distributed, using the book allows other researchers to reproduce the results presented here. Second, *David Copperfield* is available in an unusually large number of distinct editions. By 1850 Dickens was an established novelist and his works enjoyed high demand around the world. Due to this demand, many publishers issued editions of the novel. Having a variety of distinct sequences is important here because estimating profile HMMs, as they are encountered in biological sequence analysis, traditionally use hundreds of distinct homologous sequences [8]. Having the option of using many distinct sequences facilitates research exploring how varying the number of sequences used influences parameter estimates.

In the experiment I use character sequences from seven distinct copies of *David Copperfield* (Table 1). These seven sequences are associated with six editions of *David Copperfield*. The two sequences from first edition copies come from distinct items located in different university libraries. In all cases I use the machine-transcribed plain text sequences obtained from the digital library that hosts images of the item's pages (Internet Archive or Hathi Trust). No particular strategy was used for selecting items beyond checking that pages were clean.

| Year | Location | Publisher | # Characters | Repository | Identifier | Short Name |
| --- | --- | --- | --- | --- | --- | --- |
| 1850 | London | Bradbury & Evans | 17,168 | Internet Archive | personalhistoryo001850dick | 1850-LON-1 |
| 1850 | London | Bradbury & Evans | 17,459 | HathiTrust | dul1.ark:/13960/t8pc47v32 | 1850-LON-2 |
| 1858 | London | Chapman & Hall | 17,081 | Internet Archive | personalhistoryo00dickiala | 1858-LON |
| 1866 | London | Chapman & Hall | 16,618 | HathiTrust | coo.31924013471416 | 1866-LON |
| 1867 | Boston | Ticknor & Fields | 17,382 | Internet Archive | aan4786.0001.001.umich.edu | 1867-BOS |
| 1880 | Philadelphia | T. B. Peterson | 17,125 | Internet Archive | davidcopperfield01dicke | 1880-PHL |
| 1910 | Toronto | Musson Book Co. | 17,031 | Internet Archive | personalhistoryo00dickuoft | 1910-TOR |

Table 1. Transcribed character sequences from Chapter 29 of *David Copperfield*. The 1850 Bradbury & Evans edition is the first edition of *David Copperfield*. A character is a Unicode code point. Identifiers are repository-specific unique identifiers.

|  | 1850-LON-1 | 1850-LON-2 | 1858-LON | 1866-LON | 1867-BOS | 1880-PHL | 1910-TOR |
| --- | --- | --- | --- | --- | --- | --- | --- |
| 1850-LON-1 | 100% |  |  |  |  |  |  |
| 1850-LON-2 | 97.0% | 100% |  |  |  |  |  |
| 1858-LON | 94.0% | 92.6% | 100% |  |  |  |  |
| 1866-LON | 90.9% | 89.6% | 91.3% | 100% |  |  |  |
| 1867-BOS | 91.8% | 90.8% | 92.0% | 89.7% | 100% |  |  |
| 1880-PHL | 93.5% | 92.1% | 94.0% | 90.9% | 91.8% | 100% |  |
| 1910-TOR | 93.1% | 91.6% | 94.0% | 91.3% | 91.4% | 92.9% | 100% |

Table 2. Pairwise sequence identity for the seven transcribed character sequences. In calculating the percentages, the numerator is the number of identical characters and the denominator is the length in characters of the aligned sequence. Optimal alignments are found using the Needleman-Wunsch algorithm.

Table 2 shows measures of sequence identity for each pair of print sequences. Although every sequence is distinct, sequences tend to use the same characters in at least 90% of positions. Differences tend to involve the insertion or deletion of whitespace, typically due to edition-specific pagination or use of different transcription software. For example, although the aligned 1850 first edition sequences ("1850-LON-1" and "1850-LON-2") differ in 528 positions, 369 (70%) of these differences involve whitespace characters. Ignoring these positions, the percentage of matching characters increases to 99.1%. Remaining differences tend to involve different (machine) judgments about how characters should be transcribed (e.g., Rosa Dartle vs. Eosa Dartle). (Recall that virtually any "mistake" could be a faithful transcription of a printing error.) The differences between the least similar pair ("1850-LON-2" and "1866-LON") are equally superficial (see Figure 3). The number of mismatches involving whitespace is considerably greater with this pair because the editions employ different pagination. Different pagination adds to the count of character mismatches because running headers and line breaks appear in different positions.

What is notable here is that, with the exception of the two first edition sequences, the differences between pairs are idiosyncratic. No two sequences use the same pagination: running headers, line breaks, and page breaks occur in different positions. This is visible in the pairwise similarity statistics in Table 2: sequence identity is high but not too high—it is consistently below 95%. As differences in judgments about the transcription of non-whitespace characters remain rare, the differences are primarily due to edition-specific pagination.

```
All day, she seemed to pervade the whole house. If I talked to ␊ ␊␊␊304 THE PER
All day, she seemed to pervade the whole house. If I␊talked to◌◌◌◌◌◌◌◌◌◌◌◌◌◌◌◌◌

SONAL HISTORY AND EXPERIENCE ␊ ␊ Steerforth in his room, I heard her dress rust
◌◌◌◌◌◌◌◌◌◌◌◌◌◌◌◌◌◌◌◌◌◌◌◌◌◌◌◌ ◌◌◌◌Steerforth in his room, I heard her dress rust

le in ◌the little gallery outside. ␊ When he and I engaged in some of our old e
le in◌␊the little gallery outside. ◌◌When he and I engaged in some␊of our old e

 xercises on the lawn behind ␊ the house, I saw her ◌face pass from window to
 xercises on the lawn behind ◌◌the house, I saw her◌␊face pass from window to

window, like a wandering ␊ light, until it fixed itself in one, and watched us.
window, like a wandering ◌◌light,␊until it fixed itself in one, and watched us.
```

Fig. 3. Alignment of "1850-LON-2" with "1866-LON" (excerpt). Lines in the figure are wrapped for display purposes. "◌" is used as the "gap" character as the characters typically used in bioinformatics ("-" and ".") would be indistinguishable from observed characters. Control characters are replaced with Unicode replacement characters to facilitate display on the printed page. For example, "␊" replaces every line feed (line break).

### 4.2 A reference ebook sequence of the first edition of *David Copperfield*

Dickens' publisher, Bradbury & Evans, did not publish an ebook version alongside the print version of *David Copperfield* in 1850. Here I describe the construction of an ebook that approximates what Bradbury & Evans would have published had the practice of publishing ebook editions existed in 1850. This reference ebook sequence serves as the benchmark or "gold standard" against which an estimated ebook sequence will be judged.

Creating such a reference ebook edition of *David Copperfield* is made easy by the fact that ebook editions of *David Copperfield* already exist that have been painstakingly prepared by established publishers and imprints such as Penguin Classics, Oxford World's Classics, and Project Gutenberg. Given that these ebook editions are intended for use as substitutes for the first edition, they give us considerable information about what characters should appear in the reference ebook sequence. Starting from any one of these ebooks, the only step required to arrive at the reference ebook sequence is to verify that the characters in the ebook sequence have corresponding characters in the 1850 first edition. With page images of the first edition in hand, doing this for a single chapter is not onerous. (As we are only estimating a single chapter, Chapter 29, the reference ebook need only contain that chapter.) Typically only a handful of changes are required, as the following examples illustrate. The Project Gutenberg ebook neglects to mark italicized words. The Penguin ebook uses a house style with punctuation rules not used by the first edition. The Oxford ebook uses "anything" and "everything" where the first edition clearly uses "any thing" and "every thing".

To construct the reference ebook I begin with the Project Gutenberg edition and make edits such that the ebook sequence aligns with the 1850 first edition.

A comparison of the reference ebook sequence with the other ebook sequences reveals that the reference ebook sequence is essentially identical to the Oxford ebook sequence. (Inspection shows that Oxford's print and ebook editions hew close to the first edition.) Ignoring differences due to HTML tags, there are only 10 character mismatches between the reference ebook sequence and the Oxford ebook sequence. The majority of these mismatches are due to transcription errors in the Oxford ebook. (Remarkably, the Oxford ebook contains errors not present in Oxford's print version.) As there are 16,344 characters in the pairwise alignment, this means that 99.94% of characters match in the alignment between the reference ebook and the Oxford ebook.

Although the preceding comparison should allay concerns about the suitability of the reference ebook sequence (which accompanies this paper), it is possible to disagree about the component characters of any reference ebook sequence constructed for this task. Transcription of horizontal spacing, in particular, is underdetermined. There is generous spacing around quotation marks and em-dashes in the 1850 first edition—far more, in fact, than one would anticipate in a book published today. For this reason, inspecting page images will not resolve disagreements. But these disagreements will be limited to whitespace characters. Hence, alternative reference ebook sequences should be extremely similar to the one used here.

### 4.3 Evaluation

To measure the procedure's ability to estimate the desired ebook sequence, I calculate the pairwise sequence identity between the fitted profile HMM's modal sequence and the reference ebook.

Once an optimized implementation emerges for print sequences, a more nuanced evaluation should be used. Because a profile HMM often assigns many sequences roughly the same probability, gauging a model's predictive capacity using the modal sequence alone is typically regarded as inadequate [4, p. 92].

## 5 RESULTS

Table 3 shows measurements of pairwise sequence identity between the reference ebook sequence and other sequences, including the modal sequence. The third column of Table 3 reports pairwise sequence identity ignoring HTML tags and near misses. A *near miss* is a mismatch that is not semantically consequential involving an inserted space, typesetter's apostrophe, or typographic ("curly") quotation marks. For example, a mismatch between a typewriter apostrophe ' (U+0027) and a typesetter's apostrophe ' (U+2019) is a near miss. So is the insertion of a space (U+0020) in one of the sequences. Such an insertion might occur if one of the sequences uses two spaces to separate sentences and the other uses a single space. A mismatch is a mismatch, of course, and deserves to be counted as such. The first column in Table 3 reports pairwise sequence identity without any adjustments.

| Short name | % match | % match ignoring HTML tags | % match ignoring HTML tags and near misses |
|---|---|---|---|
| Profile HMM mode | 95.45 | 97.11 | 99.53 |
| Oxford ebook | 99.63 | 99.94 | 99.96 |
| Penguin ebook | 97.87 | 98.05 | 99.13 |
| PG ebook | 98.26 | 98.70 | 99.64 |
| 1850-LON-1 | 92.67 | 94.23 | 96.83 |
| 1850-LON-2 | 91.34 | 93.41 | 97.24 |
| 1858-LON | 93.04 | 94.68 | 96.87 |
| 1866-LON | 94.31 | 95.93 | 96.62 |
| 1867-BOS | 91.24 | 92.79 | 95.28 |
| 1880-PHL | 92.68 | 94.26 | 96.62 |
| 1910-TOR | 93.37 | 94.97 | 95.74 |

Table 3. Pairwise sequence identity with the reference ebook sequence. A *near miss* is a mismatch that is not semantically consequential involving an inserted space, typesetter's apostrophe, or typographic ("curly") quotation marks.

When considering raw mismatch statistics (vs. the reference ebook) it is essential to ask what characters are involved. A mismatch involving two letters tends to be more consequential for a reader's experience than a mismatch involving an extra space inserted after a period. Even letter mismatches can vary in their impact. Counting all mismatches as equivalent means failing to distinguish between, say, a mismatch between "A" (U+0061) and "ᴀ" (U+1D00, small capital A) and a mismatch between an "R" and an "E"—one print sequence features a character named "Eosa Dartle" where the six other narratives involve "Rosa Dartle".

The most important thing to note when comparing the modal sequence with the reference ebook sequence is that, with one exception, no mismatches involve letters. The mismatches all involve punctuation (quotation marks, em-dashes, commas). And the single exception is telling. It involves a correction to the 1850 first edition that appears in a majority of the print sequences from later editions.[3] The profile HMM behaves as expected, and regards the 1850 print sequences as containing deletions relative to the "consensus" sequence of *match* states. So the modal sequence contains the correction, contributing 9 mismatched characters in the alignment with the reference ebook sequence.

A visible shortcoming of the method when used on plain text print transcriptions (rather than transcriptions that attempt to mark paragraph boundaries) is that the profile HMM will occasionally preserve a line break if a majority of print sequences use an intra-paragraph line break at the same position. Across hundreds of lines, this can happen by chance. It happens 16 times in the modal sequence. Had the modal sequence used a greater variety of print sequences or used print sequences that mark paragraph breaks (e.g., transcriptions using hOCR), these 16 mismatches would not have occurred. If these 16 mismatches are excluded—and HTML tags and near-misses ignored—the pairwise sequence identity between the modal sequence and the reference ebook is essentially the same as that between the Project Gutenberg ebook and the reference ebook (60 vs. 58 mismatches out of ca. 16,100 characters). Given that the Project Gutenberg edition required many hours of human labor to prepare and proofread, this result is remarkable.

## 6 LIMITATIONS

The chief limitation is that the method requires at least two print editions with distinct pagination. Other limitations involve HTML tags that are present in the reference ebook but are missing in the modal sequence. Not only are these HTML tags missing in the estimated ebook version, the fitted profile HMM will assign lower probability to an ebook sequence with these HTML tags than to one without them. Fortunately, these limitations are either easy to resolve or of bounded consequence.

The main limitation of the method, that it requires at least two distinct editions, does not render the method useless. Books that were popular around or after the time they were published frequently have at least two editions. (*David Copperfield* likely has hundreds.) Most of these (once) popular books lack editions in an accessible format. So the procedure using a profile HMM can still facilitate the creation of many accessible ebooks from print copies.

The failure of the profile HMM to assign higher probability to ebook sequences with the correct HTML tags (e.g., <p>, <em>, <h2>) has a ready solution: use a more descriptive (machine) transcription. Using a transcription of the printed page that includes additional information about line spacing and typeface changes would solve the problem of missing paragraph, section heading, and emphasis tags. For example, had the experiment described above used a more detailed transcription instead of plain text, paragraph tags would already have been present in the print sequences because the relevant format (hOCR) uses XHTML, which marks paragraph boundaries in the same manner as the reference ebook sequence. So the profile HMM would have included three match states at the beginning of each paragraph capturing the observed regularity of the three-character sequence (<, p, >) tending to occur in the same position across several editions. Using a more detailed transcription would also resolve the issue of absent section heading and emphasis tags as changes in font size and shape are typically noted in detailed (machine) transcriptions.

---

[3]The first edition has "As she still looked fixedly…" whereas later print sequences have "As she still stood looking fixedly.…". The word "stood" is inserted and "looked" is changed to "looking". The edit involves a sentence appearing on page 305 of the first edition.

Illustrations, figures, and bespoke glyphs pose no theoretical problem for an approach using a profile HMM because these elements of the text can be "transcribed" using PETSCII, SVG, or embedded using a data URI. As a practical matter, however, each such transcription is likely to be unique and will never be associated with a *match* state—only with an *insert* state. Some sort of special handling is likely to be required.

## 7 DISCUSSION

This paper offers an eye-catching demonstration of one potential use of a profile HMM. Due to space constraints, I can only gesture at the myriad potential uses of profile HMMs in the study of transcribed sequences of books and periodicals. Since their introduction in 1993, profile HMMs have been used in numerous different tasks in biology [5]. For every task in biology, it is easy to imagine an analogous task of interest to researchers in information retrieval, book history, and media studies. For example, the task profile HMMs initially addressed was *search* [8, 11]. Profile HMMs proved capable of identifying closely- and distantly-related proteins and RNAs. Translated into the context of information retrieval, profile HMMs used in this way would identify (inexact) copying of particular paragraphs in other works. One setting where this would be extremely useful is the study of works with variant editions (e.g., David Mitchell's *Cloud Atlas* [7]). Profile HMMs have also been used to infer plausible phylogenies, a task familiar to researchers studying written documents [1] that is also of interest to those attempting to automatically infer publication dates for the countless books, pamphlets, and articles that lack reliable publication information.

A full appreciation of the use of profile HMMs in biology and their potential uses in the study of text documents would begin by abandoning the distinction between print and ebook sequences. The profile HMM, after all, knows nothing about such a distinction. A more productive and intellectually satisfying use of a profile HMM in the case of books would be to model all homologous sequences together. An ebook sequence is obviously a homologous sequence. So too are versions of a work composed using SGML, XML-TEI, Markdown and LaTeX. (The distinction is also unverifiable: an ebook sequence can be made into a print sequence by printing it out and then using OCR; any print sequence can be made into an ebook sequence by saving it to a file and declaring it a digital edition.) A profile HMM can and should be used to model all sequences in a family.[4]

## 8 CONCLUSION

A profile hidden Markov model of character sequences can be used to estimate an ebook edition from several print copies of a given work. The resulting ebook has the desired properties we anticipate in a publisher-prepared ebook. This technique is successful because the ebook edition features only those elements shared by a majority of the print copies. The profile HMM models these shared elements using *match* states. The model uses *insert* states to model elements found in only one or two print editions, elements such as running headers, page numbers, end-of-line hyphenation, and page breaks. As the profile HMM assigns higher probability to sequences that tend to feature only shared elements, high probability sequences under the model tend to resemble an ebook sequence. The paper verifies this through an experiment using seven copies of a nineteenth-century novel.

The availability of this method has benefits for libraries and readers, in particular readers with print disabilities who require books in an accessible format.

---

[4]When estimating a profile HMM using many sequences that are extremely similar—derived, say, from the same item using different OCR software—care needs to be taken to appropriately weight training sequences. For a standard solution to the problem in biological sequence analysis see Durbin et al. [4, §5.8].

**ACKNOWLEDGEMENTS**

Thanks in particular to Christof Schöch, whose work on text comparison [14] prompted the search that led me to learn about the profile HMM. Many thanks as well to Dan Rockmore, Yohei Igarashi, Troy Bassett, Michael Betancourt, and Debora Shaw for valuable feedback on drafts of the paper.

## A   INITIAL PARAGRAPHS OF THE PROFILE HMM'S MODAL SEQUENCE

The following shows the first 12 lines of the modal sequence from a profile HMM estimated using seven copies of *David Copperfield*. Lines have been wrapped at 72 columns so they will fit on these pages. Line breaks due to wrapping are preceded by "↵". Other line breaks (without a visible symbol) indicate the presence of a line feed character (U+000A) in the sequence itself.

   In the first edition there is a page break with accompanying page number (p. 304) and a running header immediately after "If I talked to" (part of the final paragraph shown below).

```
CHAPTER XXIX.

I VISIT STEERFORTH AT HIS HOME, AGAIN.

I MENTIONED to Mr. Spenlow in the morning, that I wanted leave of ↵
absence for a short time ; and as I was not in the receipt of any ↵
salary, and consequently was not obnoxious to the implacable Jorkins, ↵
there was no difficulty about it. I took that opportunity, with my voice ↵
sticking in my throat, and my sight failing as I uttered the words, to ↵
express my hope that Miss Spenlow was quite well ; to which Mr. Spenlow ↵
replied, with no more emotion than if he had been speaking of an ↵
ordinary human being, that he was much obliged to me, and she was very ↵
well.

We articled clerks, as germs of the patrician order of proctors, were ↵
treated with so much consideration, that I was almost my own master at ↵
all times. As I did not care, however, to get to Highgate before one or ↵
two o'clock in the day, and as we had another little excommunication ↵
case in court that morning, which was called The office of the Judge ↵
promoted by Tipkins against Bullock for his soul's correction, I passed ↵
an hour or two in attendance on it with Mr. Spenlow very agreeably. It ↵
arose out of a scuffle between two churchwardens, one of whom was ↵
alleged to have pushed the other against a pump ; the handle of which ↵
pump projecting into a school-house, which school-house was under a ↵
gable of the church-roof, made the push an ecclesiastical offence. It ↵
was an amusing case ; and sent me up to Highgate, on the box of the ↵
stage-coach, thinking about the Commons, and what Mr. Spenlow had said ↵
about touching the Commons and bringing down the country.

Mrs. Steerforth was pleased to see me, and so was Rosa Dartle. I was ↵
agreeably surprised to find that Littimer was not there, and that we ↵
were attended by a modest little parlor-maid, with blue ribbons in her ↵
cap, whose eye it was much more pleasant, and much less disconcerting, ↵
to catch by accident, than the eye of that respectable man. But what I ↵
particularly observed, before I had been half-an-hour in the house, was ↵
the close and attentive watch Miss Dartle kept upon me ; and the lurking ↵
```

manner in which she seemed to compare my face with Steerforth's, and Steerforth's with mine, and to lie in wait for something to come out between the two. So surely as I looked towards her, did I see that eager visage, with its gaunt black eyes and searching brow, intent on mine ; or passing suddenly from mine to Steerforth's ; or comprehending both of us at once. In this lynx-like scrutiny she was so far from faltering when she saw I observed it, that at such a time she only fixed her piercing look upon me with a more intent expression still. Blameless as I was, and knew that I was, in reference to any wrong she could possibly suspect me of, I shrunk before her strange eyes, quite unable to endure their hungry lustre.

All day, she seemed to pervade the whole house. If I talked to Steerforth in his room, I heard her dress rustle in the little gallery outside. When he and I engaged in some of our old exercises on the lawn behind the house, I saw her face pass from window to window, like a wandering light, until it fixed itself in one, and watched us. When we all four went out walking in the afternoon, she closed her thin hand on my arm like a spring, to keep me back, while Steerforth and his mother went on out of hearing : and then spoke to me.